\documentclass[%
 reprint,
 amsmath,amssymb,
 aps,
]{revtex4-2}

\usepackage{graphicx}
\usepackage{dcolumn}
\usepackage{bm}
\usepackage{caption}
\usepackage{subcaption}
\usepackage{graphicx}
\usepackage{calrsfs}
\usepackage[labelformat=parens,labelsep=quad, skip=3pt]{caption}
\DeclareUnicodeCharacter{00A0}{ }
\usepackage[utf8x]{inputenc}

\begin{document}

\preprint{APS/123-QED}

\title{Disassortative mixing of boundedly-rational players in socio-ecological systems}



\author{Prasan Ratnayake}
\email{pshanarat@gmail.com}
\affiliation{
 Department of Physics, Faculty of Science,\\
 University of Colombo,\\
 Colombo, Sri Lanka.
}%

\author{Dharshana Kasthurirathna}
\affiliation{%
 Faculty of Computing,Sri Lanka
 Institute of Information Technology,B263,Malabe 10115,Sri Lanka.
}%
\email{dharshana.k@sliit.lk}

\author{Mahendra Piraveenan}
\affiliation{%
 Complex Systems Research Group, Faculty of Engineering, University of Sydney, Camperdown NSW 2006, Australia.
}%
\email{mahendrarajah.piraveenan@sydney.edu.au}



\date{\today}

\begin{abstract}
\begin{description}
\item[Abstract]
Bounded rationality refers to the non-optimal rationality of players in non-cooperative games. In a networked game, the bounded rationality of players may be heterogeneous and spatially distributed. It has been shown that the `system rationality', which indicates the overall rationality of a network of players, may play a key role in the emergence of scale-free or core-periphery topologies in real-world networks. On the other hand, scalar-assortativity is a metric used to quantify the assortative mixing of nodes with respect to a given scalar attribute. In this work, we observe the effect of node rationality-based scalar-assortativity, on the system rationality of a network. Based on simulation results, we show that irrespective of the placement of nodes with higher rationality, it is the disassortative mixing of node rationality that helps to maximize system rationality in a population. The findings may have useful interpretations and applications in socio-economic systems in maximizing the utility of interactions in a population of strategic players.\\

\item[Keywords]\textbf{Bounded rationality;Assortativity;Evolutionary games;Network science}
\end{description}
\end{abstract}

\maketitle

\section{Introduction}

The relationship between non-optimal bounded rationality of networked players and network topology has been explored in the field of networked games. Recent work by Kasthurirathna et al~\cite{kasthurirathna2015emergence} quantified the `system rationality' of a networked system as the negative average of Jensen-Shannon divergence between the Nash equilibria and the Quantal Response Equilibria (QRE) of all games played in a single iteration of a networked game. The rationale for this definition was that the more a population diverges from Nash equilibria, the more irrational it is, on average. Therefore, perfect system rationality is indicated by a value of zero and the more negative the system rationality value is, the more irrational the system would be.  Kasthururathna et al~\cite{kasthurirathna2015emergence} demonstrated that when a population of random networks with the same and constrained average rationality goes through evolutionary process, where interactions with higher `system rationality' were preferentially selected, scale-free networks emerge as a result. Therefore, they postulated that evolutionary pressure to increase the `system rationality' may have been a reason why scale-free networks emerged in socio-ecological systems. 

It is important to note that while scale-free networks emerge from random networks when evolutionary pressure is applied to optimise system rationality, this does not imply that scale-free topology is the optimal topology to maximise system rationality. Roman and Brede~\cite{roman2017topology} argued that core-periphery (or hub-and-spoke) network structures\cite{rombach2014core} are optimal for system rationality maximisation.  Meanwhile, Law et al~\cite{law2019placement} has argued that, public utility of a networked system depends on placement of players, and scale-freeness and disassortativity in networks are positively correlated to the public utility. However, their observation was about public utility, not system rationality, and these, though could be correlated, are distinct concepts.  


The existing work on this topic is predominantly based on the presumption that the degree of nodes positively correlates with node rationality. It is based on this premise that scale-free and core-periphery network topologies have been argued to maximize system rationality or public utility, respectively. Extending from this body of work, here we argue that it's the assortative mixing of node rationality that plays a prominent role in determining system rationality, and not necessarily the placement of the nodes with higher rationality.  

To do so, we observe how the assortative mixing of the bounded rationality of players affect the system rationality. Thereby, we show that it is immaterial whether the `smartest' or more rational players are assigned to the hubs, as done in Kasthurirathna et al~\cite{kasthurirathna2015emergence} and Roman and Brede~\cite{roman2017topology} or it is the least rational players that are assigned to the hubs, as long as the node rationality-based  scalar disassortativity of the network is maximised. We show that if topologies can be re-wired, then networks with the highest disassortativity may optimise the system rationality, and the results obtained by Roman and Brede~\cite{roman2017topology} in terms of core-periphery configurations is a special case of this. 

The rest of this paper is organised as follows: in section \ref{s-Background}, we provide the background and discuss relevant literature that motivated this study. In section \ref{s-Definitions}, we define the key concepts that are used in our analysis. In section \ref{s-Results}, we describe our simulation experiments and our results. Finally, in section \ref{s-Conclusions} we discuss our results, summarise our conclusions and point some directions for future work.

\section{Background} \label{s-Background}

Network science attempts to model and predict the structure and operations of entities that are connected over non-trivial topologies\cite{newman2003structure,albert2002statistical}. It has applications in myriad fields such as epidemiology, social network analysis and biology. The prevalence of the scale-free topology has been observed in many real-world networks ranging from online social networks to neural networks in the brain\cite{newman2003structure,chung2014topology}. 

Autonomous agents distributed over a topology may interact with their neighbors in making strategic decisions. These strategic interactions may be modelled as strategic games in game theory, where each agent acts in a self-interested manner to maximize its payoff. Strategic games that are played over a population of players that interact over a topology can be used to model biological and socio-economic systems. Evolutionary game theory utilizes such networked games to model the evolution of strategies over a network of players\cite{weibull1997evolutionary}.

Evolutionary stable strategies refer to a dynamic equilibrium of a strategy that may overrun the competing strategies\cite{taylor1978evolutionary}. It has been shown that the topology of a network is critical in determining the evolutionary stability of a strategy, as well as the general evolution of strategies over time. Well known iterative games such as the prisoner's dilemma game and the stag-hunt game are often applied on networks of players to simulate the evolution of strategies\cite{santos2006graph,kasthurirathna2015evolutionary,kasthurirathna2014influence}.  


Evolution of cooperation in networked games has been shown to be dependent on the network topology, where scale-free networks have been shown to facilitate cooperation among agents. On the other hand, individual intrinsic characteristics such as the cognitive capacity and intelligence have been shown to have an effect on the evolution of cooperation in populations\cite{devaine2014theory,han2011intention,mcnally2012cooperation}. In this work, we do not focus specifically on the evolution of cooperation, but the effect assortative mixing of node rationality on facilitating more rational interactions with respect to a given strategic decision making scenario, where mutual cooperation may be one such rational strategy.  



The effect of assortative mixing on the evolution of cooperation has been explored previously\cite{rong2007roles,rong2009effect}. It has been shown that in networked prisoner's dilemma and public goods games, degree based diassortativity facilites cooperation among players. This is due to the tendency of cooperative hubs to be protected from other defective hubs in a disassortative network. However, the existing work on the evolution of cooperation assumes that all nodes have optimal rationality in a network of players, which is a very strong assumption. 


\subsection{Prisoner's dilemma game}
\begin{figure}
{
\centering\includegraphics[width=5.5cm]{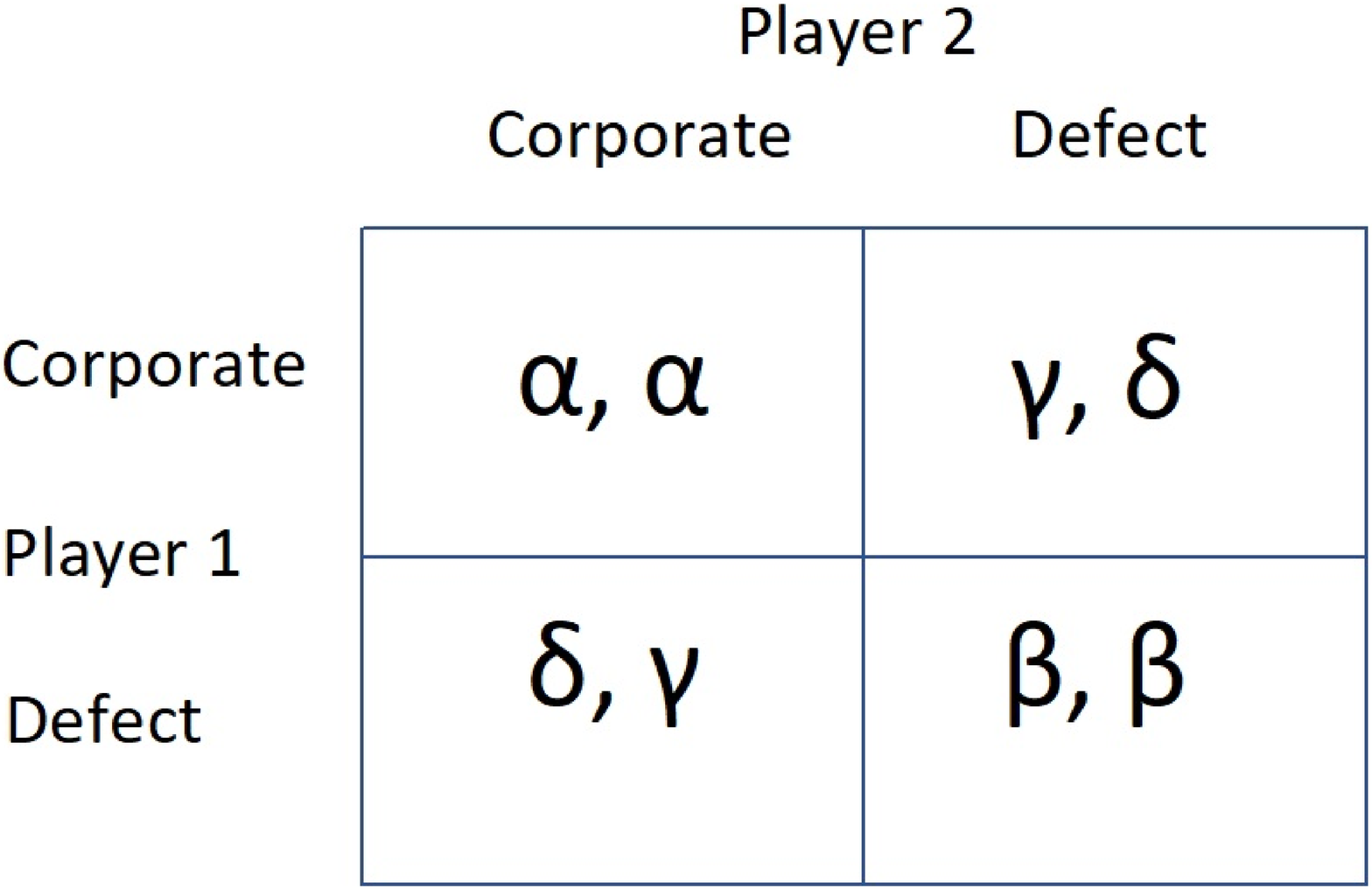}
}
\caption{Payoff matrix of the prisoner's dilemma game}
\label{game}
\end{figure}

Prisoner's dilemma game\cite{axelrod1980effective} is one of the most commonly used games in networked games to model social interaction. In this work, the prisoner's dilemma game payoff matrix denoted by Fig.\ref{game} is used for modeling social interactions among neighboring players. In prisoner's dilemma game, the payoffs satisfy the inequality conditions $\delta>\alpha>\beta>\gamma$. 

\subsection{Bounded Rationality}

Nash Equilibrium is one of the cornerstones in game theory that theorizes that in every strategic interaction, there exists one or more equilibria, where no player would benefit deviating from \cite{myerson1999nash}. However, most real-world strategic interactions deviate from Nash equilibrium since Nash equilibrium assumes perfect or optimal rationality. Real-world players possess non-optimal or bounded rationality, as they may have limited information, time and cognitive capacity to chose a strategy\cite{conlisk1996bounded}. This bounded rationality of players can be quantified using a rationality parameter in the quantal response equilibrium (QRE) model\cite{mckelvey1995quantal}. In the QRE model, $\lambda_{i}$ denotes the individual rationality of player $i$. The rationality parameter enables controlling the ability of player $i$ to respond to the strategy distribution of the opponent and the payoffs that correspond to the respective strategies. In other words, the rationality parameter makes it possible to adjust player $i$'s bounded rationality value. When $\lambda_{i} \rightarrow \infty$, the quantal response equilibrium reaches the corresponding Nash equilibrium, and as $\lambda_{i} \rightarrow 0$, the player would operate in a totally random, thus irrational, manner\cite{goeree2010quantal}. Thus, the logit function used in the QRE model provides a convenient mechanism to derive the mixed strategy equilibrium probability distributions of strategic agents that may possess non-optimal rationality values.  

As players in a population may possess the factors contributing bounded rationality in varying degrees, the bounded rationality of a population may be heterogeneous\cite{kasthurirathna2015emergence}. Thus, it is necessary to account for the heterogeneity of the bounded rationality of players to better fit the game theoretic models to real-world strategic interactions.





 Even though there have been numerous attempts to quantify the bounded rationality of players, they have predominantly focused on quantifying the bounded rationality as a population specific uniform value that is applied to all players that operate in a given strategic decision making environment, in a generic manner. For instance, Wolpert~\cite{wolpert2006information} proposed a method to analytically quantify the bounded rationality of a given player by solving the Maxent or maximum entropy Lagrangians that uses the Boltzman distribution to derive the mixed strategy probability distribution of a human agent. However, empirical evidence as well as theoretical frameworks such as the cognitive hierarchical model~\cite{camerer2004cognitive} strongly suggest that the bounded and non-optimal rationality of agents in a given population is typically distributed in a heterogeneous fashion and not as a population specific uniform value. Numerous reasons such as the variability of the cognitive capacity of nodes and the varying degrees of information available to nodes due to the different levels of social interaction, may lead to this heterogeneity in bounded rationality.

There have been numerous prior attempts made to quantify the heterogeneous rationality of players using the quantal response equilibrium model~\cite{rogers2009heterogeneous,golman2012homogeneity}. Examples of such models include the heterogeneous QRE model and the Truncated QRE model~\cite{rogers2009heterogeneous}. Further, the cognitive hierarchical model~\cite{camerer2004cognitive} that operates under the same goal can be considered as a special case of Truncated QRE model. However, such models operate by varying the player rationality $\lambda_i$, in such a way that it fits with the empirical results in a given setting. The key limitation in such an approach, in our opinion, is that the rationality parameter is being used as an arbitrary parameter, without being based on any measurable real-world attribute of an agent. Even though this may improve the adaptability of the rationality parameter, it may limit the real-world applicability of such models, in predicting the outcome of interactions among strategic players with non-optimal rationality.   

\subsection{Modelling bounded rationality distributions}

Several recent studies have considered how a rationality distribution could be modelled with respect to a given network topology~\cite{kasthurirathna2015emergence,roman2017topology,roman2018dynamic,kasthurirathna2016modeling}. The number of neighbors that an agent in a population has, which is commonly referred to as the degree of a node in network analysis, may indicate the amount of social interactions that the agent may have within the population in concern. However, there may be other factors that affect the level of social interaction, which are known as the `tie strength' attributes. Examples of such attributes may be the volume of information exchanged or even the frequency of interactions between a given pair of players. Moreover, there could be either a linear or a non-linear correlation between the social interaction an agent has an the agent's bounded rationality. Recent studies have attempted to model such correlations ~\cite{kasthurirathna2015emergence,kasthurirathna2016modeling}, where the bounded rationality of an agent is quantified based on its weighted degree or the degree in the case of an unweighted network, as shown in Eq. \ref{eq8}.

\begin{equation} \label{eq8}
\lambda_{i} = r.f(\sum_{j=1}^{n}w_{ij})
\end{equation}

Here, $\lambda_{i}$ is the bounded rationality of an agent $i$; $r$ is referred to as the `network rationality parameter'. It is a property of the network or the population that captures the sensitivity of the topological features to the rationality of an agent. In other words, the average rationality of a given population,  $\bar\lambda$, is proportional to network rationality parameter. This is due to the fact that any deviation in $r$ will cause a corresponding shift in the bounded rationality value $\lambda_{i}$ of an agent $i$ in the population. Further, $w_{ij}$ represents the weight of the link connecting an agent $i$ with each neighbour $j$, where $n$ is the number of neighbours that an agent $i$ has. Accordingly, an agent may operate completely randomly if the network rationality parameter is set to $r=0$ or if the degree of an agent is zero. On the other hand, an interaction may reach that of Nash equilibrium if the network rationality parameter $r\rightarrow\infty$, or when the weighted degrees of respective nodes are very large values. 
The approach used in the existing literature in modelling bounded rationality as a function of degree may be considered a special case of a more generalized model that is used in this work, which is based on an assortativity-based distribution of bounded rationality values.
 

\subsection{Bounded rationality in networked games}

One of the key aspects of evolutionary games is the effect that the payoffs of neighboring agents have on the strategy adoption of an agent. Moreover, the rationality of agents may also be critical in the payoffs obtained by each agent, as most real-world players or agents may make non-optimal decisions.  

There has been some recent interest in the topological effect on the rationality of players in a network and vice versa. For instance, it has been shown that certain topologies such as scale-free topology and core-periphery topology facilitate more rational interactions than the other topologies. Kasthurirathna et al. \cite{kasthurirathna2015emergence} have shown that scale-free topologies emerge based on the premise that the degree of nodes infer the rationality values and rationality of a node is a defined as a function of degree. Roman and Brede \cite{roman2017topology} showed that under the same assumptions, core-periphery topologies best facilitate more rational interactions, compared to other topologies. Gunewardena et al. \cite{gunawardana2019information} further extended this approach to infer the rationality of nodes based on the incoming information flows. 

The underlying assumption of topology based or information flow based rationality is that the information availability at a node to make more rational decisions can be inferred by the centrality of a node or the cumulative directed incoming information flow towards node. In both these scenarios the cognitive capacity of nodes and the time available to compute the optimal decisions are assumed to be homogeneous. 

To the best of our knowledge, existing work have not specifically looked at the effect of mixing patterns of rationality among players on the system rationality of a population. In this work, we attempt to decouple the distribution of bounded rationality from the underlying topology or the directed information flows and observe the effect of the distribution of rationality and the mixing of rationality among players on the overall rationality of population. 

\section{Definitions} \label{s-Definitions}

\subsection{The Quantal Response Equilibrium}

The  \textit{logit} function given in Eq.\ref{eq1} is used for computing the Quantal Response equilibrium ~\cite{goeree2008quantal,zhang2013quantal}.

\begin{equation}  \label{eq1}
P_{j}^{i} = \frac{e^{\lambda_{i} E^{i}(s_{j}^{i},\mathbf{P})}}{\sum\limits_{k=1}^{\kappa}e^{\lambda_{i} E^{i}(s_{k}^{i},\mathbf{P})}}
\end{equation}

Here, $P_{j}^{i}$ is the probability of player $i$ adopting the strategy $j$. $E^{i}(s_{j}^{i},\mathbf{P})$  is the expected utility of an agent $i$ in choosing strategy $j$, provided that other agents adopt their respective strategies according to the probability distribution $\mathbf{P}$. The parameter $\kappa$ represents the total number of strategies that player $i$ can choose from. The value of the rationality parameter $\lambda_{i}$ may vary from zero to positive infinity.

In a two-player prisoner's dilemma game, applying the QRE model may derive Eq.\ref{eq9} and Eq.\ref{eq10} from Eq.\ref{eq1} to represent the probabilities based on which the two agents may adopt cooperation as their strategy, based on their mixed strategy quantal response equilibrium.

\begin{equation} \label{eq9}
p^{1}_{c} = \frac{e^{\lambda _{1}(p^{2}_{c}u^1_{11} + (1-p^{2}_{c})u^1_{12})}}{e^{\lambda _{1}(p^{2}_{c}u^1_{11} + (1-p^{2}_{c})u^1_{12})} + e^{\lambda _{1}(p^{2}_{c}u^1_{21} + (1-p^{2}_{c})u^1_{22})}}
\end{equation}

\begin{equation} \label{eq10}
p^{2}_{c} = \frac{e^{\lambda _{2}(p^{1}_{c}u^2_{11} + (1-p^{1}_{c})u^2_{21})}}{e^{\lambda _{2}(p^{1}_{c}u^2_{11} + (1-p^{1}_{c})u^2_{21})} + e^{\lambda _{2}(p^{1}_{c}u^2_{12} + (1-p^{1}_{c})u^2_{22})}}
\end{equation}

Using the utility values we have chosen ($u_{11}^{1}=3, u_{11}^{2}=3, u_{12}^{1}=0,  u_{12}^{2}=5,  u_{21}^{1}=5,  u_{21}^{2}=0,  u_{22}^{1}=1,  u_{22}^{2}=1$), these can be simplified to:

\begin{equation} \label{eq9.1}
p^{1}_{c} = \frac{e^{\lambda _{1}(3p^{2}_{c})}}{e^{\lambda _{1}(3p^{2}_{c})} + e^{\lambda _{1}(4p^{2}_{c}+ 1)}}
\end{equation}

\begin{equation} \label{eq10.1}
p^{2}_{c} = \frac{e^{\lambda _{2}(3p^{1}_{c})}}{e^{\lambda _{2}(3p^{1}_{c})} + e^{\lambda _{2}(4p^{1}_{c}+ 1)}}
\end{equation}

Here, $p^{1}_{c}$ and $p^{2}_{c}$ represent the probabilities using which agent $1$ and $2$ may cooperate, respectively. Further, $\lambda_{1}$ and $\lambda_{2}$ denote the rationality values of players $1$ and $2$, respectively. In the existing work, these values are derived based on their respective degrees or weighted degrees, the rationality function used and the network rationality parameter used, as denoted in Eq.\ref{eq8}. The resulting probability distribution represent the QRE for the particular pair of agents engaging in a strategic interaction.  

The Nash equilibrium for this game is when both players defect, where $p^{1}_{c}=0$, $p^{1}_{d}=1$ and $p^{2}_{c}=0$, $p^{2}_{d}=1$. 


\subsection{The Jensen-Shannon Divergence}

The Kullback--Leibler (KL) divergence~\cite{cover1991entropy}, which is also referred to as the relative entropy, is commonly used to measure the divergence between two probability distributions. Given two probability distributions $P$ and $Q$, the Kullback--Leibler (KL) divergence of them is given by,

\begin{equation}  \label{eq3}
KL(P||Q) = \sum_{i}P_i ln\frac{P_i}{Q_i}
\end{equation}

where $P_i$ and $Q_i$ represent the two probability distributions in concern, respectively. It has to be noted that the KL divergence is an asymmetric measure, which has directionality, where the divergence from $P$ to $Q$ is not equal to the divergence from $Q$ to $P$. 

An equivalent symmetric measure that is derived using the directional KL divergence values can be obtained using the Jensen-Shannon Divergence\cite{menendez1997jensen}. Eq. \ref{eq3} denotes the definition of Jensen-Shannon Divergence. Here, $M_i$ is the probability distribution derived by averaging the two probability distributions $P$ and $Q$.

\begin{equation}  \label{eq3}
Div(P||Q) = \frac{\sum_{i}P_i ln\frac{P_i}{M_i} + \sum_{i}Q_i ln\frac{Q_i}{M_i}}{2}
\end{equation}

In this work, we use the Jensen-Shannon divergence to quantify the divergence of a QRE based equilibrium from the equivalent Nash equilibrium of a given interaction. It has been applied in previous work for the same purpose\cite{kasthurirathna2015emergence,roman2017topology}.


\subsection{Average Rationality}

The average rationality $\bar{\lambda}$ of a network is the average value of the rationality values of all nodes in the population. In a given rationality distribution, the average rationality remains independent of the variations in the mixing patterns, as long as the rationality distribution is unchanged.  

\subsection{System rationality}

We employ the average Jensen-Shannon divergence between the Quantal Response and Nash equilibria to quantify the system rationality\cite{kasthurirathna2015emergence}. The justification for this is the premise that the more self-interested and rational the agents are, the less the divergence from the corresponding Nash equilibrium will be.

To obtain a singular population specific rationality for a network players that interact based on a given topology, the averaged value of the Jensen-Shannon divergence values of all interactions over all the links of the network is measured. The system rationality $\rho$ is given by Eq. \ref{eq-ra}. Accordingly, the system rationality is inversely proportional to the averaged divergence of the interactions from the corresponding Nash equilibria. In Eq. \ref{eq-ra}, $N_K$ denotes the probability distribution given by the Nash equilibrium at link $k$. $Q_k$ denotes the probability distribution of the quantal response equilibrium at link $k$ and $M$ denotes the number of links in the given network.


\begin{equation} \label{eq-ra}
\rho({\mathcal{N}}) = -\frac{1}{M}\sum_{k=1}^{M}Div(N_k||Q_k)
\end{equation}



This average Jensen-Shannon divergence is different from the average rationality, $\bar\lambda$. $\bar\lambda$ denotes the average of the heterogeneous rationality values assigned to each agent. However, two populations with the same  $\bar\lambda$ may have different system rationality values, since it's possible to have varying concrete topologies with varying assortativity values, among populations of players with the same degree distribution.




\subsection{Assortativity and Scalar-assortativity}

Assortativity measures the tendency of nodes to connect with other nodes that are similar to them based on a particular attribute, or the `assortative mixing', of a given network. Degree assortativity, which is commonly referred to as the assortativity of a network, is the Pearson correlation~\cite{newman2002assortative,newman2003mixing,piraveenan2010assortative} between the expected-degree distribution $q_k$, and the joint-degree distribution $e_{j,k}$ in a given network. The expected degree distribution is the probability distribution of traversing the links of the network, and finding nodes with degree $k$ at the end of the links. On the other hand, the joint degree distribution represents the probability distribution of a link having $j$ in one node and the degree $k$ on the node at the other end. The normalized Pearson coefficient of $e_{j,k}$ and $q_k$ gives us the assortativity coefficient $P_a$, for an undirected network.

Assortativity coefficient is denoted by,

\begin{equation}
\label{eq1.9} P_a = \frac{1}{\sigma _q^2 }\left[ {(\sum \limits_{j k} {j k e_{j,k} } ) - \mu _q^2 } 
\right]
\end{equation}

where $e_{j,k}$ is the joint probability distribution of the excess degrees of the two nodes that form a random link. Here, $\mu_{q}$  and  $\sigma_{q }$ are the expected value or mean, and the standard deviation,  of the excess degree distribution $q_k$, respectively. If a network has optimal assortativity ($P_a = 1$), then all nodes will mix with the nodes with exactly the same degree. If the network has zero assortativity, that is if $P_a = 0$, then node degree has no correlation with the mixing of nodes and thus, a given node may randomly connect with any other node. If a network is perfectly disassortative ($P_a = -1$), all nodes will connect to neighbors with extremely different degrees. A good example of a perfectly disassortative network is a star network. In general, complex networks with star `motifs' in them, such as the core-periphery networks~\cite{roman2017topology} may be more disassortative.   

Assortativity may also be quantified based on any particular scalar attribute of the node. For instance, in a social network, the assortativity may be measured using scalar attributes such as ethnicity, religion, age, income etc., which may indicate the assortative mixing of nodes with respect to a given scalar attribute. 
Assortativity measured based on a scalar attribute, which is referred to as`scalar-assortativity', is given by \cite{newman2003mixing},



\begin{equation}
\label{eq1.8} P_s=\frac{1}{1 -  {\sum\limits_{jk}  {a_j
 b_k } } }\left[ {\sum\limits_{jk} {\left( {e_{j,k} -a_j
 b_k} \right)} } \right]
\end{equation}

where $a_j$ and $b_k$ are the fraction of each type of end of a link that is attached to node of type $j$ and node of type $k$ . For undirected networks, where there is no specific `source' or `target' node, $a_j= b_j$. Likewise, $e_{j,k}$ is the proportion of links which have a node of type $j$ as the source and a node of type $k$ as the target. Again, in undirected networks $e_{j,k} = e_{k,j}$.

Both degree and scalar assortativity can be used to gain insights of a structure of a network that may not be apparent in the degree distribution. The reason is that the mixing patterns can vary even in multiple network topologies with the same degree distribution. 

In this work, we use the bounded rationality of each player as the scalar attribute, using which we compute the node rationality-based scalar-assortativity of a network. This enables us to observe how the assortative mixing of bounded rationality affects the system rationality of a networked population of players.

\section{Methodology}
In order to observe how the bounded rationality based scalar-assortativity affects system rationality, we distributed the bounded rationality values in an Erdős–Rényi random network\cite{albert2002statistical} of $1000$ nodes and $3000$ links based on the following distribution functions. The size of the network was determined based on similar simulations done in previous work\cite{kasthurirathna2015emergence,roman2017topology}. Both the minimum degree and the number of connected components of the network was one, indicating that there were no isolated nodes and the network was connected.

\subsection{Bounded rationality distributions}

In this work, we use two distinct approaches for modelling the rationality values of players in a network: 1) assigning rationality values in such a way that they are positively correlated with the node degree 2) assigning rationality values in such a way that they are negatively correlated with the node degree. Using these two approaches we could observe the effect of assortative mixing patterns of node rationality on the system rationality. 

\subsubsection{Node rationality assigned to be positively correlated with node degree}

\begin{equation} \label{f1}
\lambda_{i} = r*e^{k_i}
\end{equation} where $k_i$ is the degree of node $i$.

\subsubsection{Node rationality assigned to be negatively correlated with node degree}

\begin{equation} \label{f2}
\lambda_{i} = r*e^{\frac{1}{k_i}}
\end{equation} where $k_i$ is the degree of node $i$.

In both the above functions, the $r$ value, which denotes the network rationality parameter, was set to $0.01$. This value was selected in order to obtain a broad distribution of rationality values, based on experimental results conducted using the same payoff values of the prisoner's dilemma game, which has been reported in the previous work \cite{kasthurirathna2015emergence}. 

Based on each of the given functions, the bounded rationality of each node was computed in the random network. Then, under each rationality model, two links were selected randomly and were rewired to check whether the resulting interactions improve the system rationality of the network. If the system rationality improved after the rewiring process, the new pair of links is preserved. If the aforementioned rewiring did not improve the system rationality, the rewiring was reverted and the previous pair of links was re-established. This method of optimizing the system rationality was inspired by simulated annealing optimization technique\cite{van1987simulated} and has been previously used to observe the topological evolution when system rationality is incrementally improved\cite{kasthurirathna2015emergence}.

After each iteration, the bounded rationality of each node was recalculated based on the functions given in Eq. \ref{f1} and Eq. \ref{f2}, depending on the scenario being tested. This process was repeated for $2000$ iterations until the system rationality values were stabilized and each intermediate network was stored to measure their network properties.  

The scale-free exponent\cite{albert2002statistical,barabasi1999emergence}, scale-free correlation\cite{barabasi1999emergence,albert2002statistical}, scalar-assortativity and the system rationality of each intermediary network were computed, for both the scenarios where the node rationality was assigned to be positively and negatively correlated with the node degree. The variation of the scale-free exponent and scale-free correlation gives an indication on the evolution of the topology through the rewiring process to improve the system rationality. 

In addition, the system rationality and scalar-assortavitity based on the node rationality were computed in each intermediary network. The variation of scalar-assortativity and the system rationality signifies effect of the assortative mixing of node rationality on the system rationality. These network properties of the intermediary networks were averaged over $20$ such repeated experiments, in order to normalize the the effect of randomness in rewiring the links to incrementally improve system rationality. The resulting values were plotted against the number of rewiring iterations to observe how the evolution of the network topology affects the scale-free exponent, scale-free correlation, system rationality and scalar-assortativity. In the following section, we present the results obtained from the simulations described in this section.  








\section{Results} \label{s-Results}


 \begin{figure*}
 \centering
 \begin{subfigure}{8cm}
 \centering\includegraphics[width=8cm]{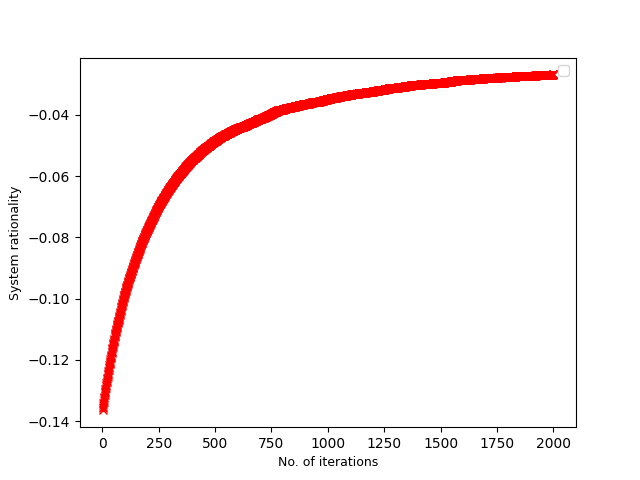}
 \caption{Node rationality positively \newline correlated with degree}
 \end{subfigure}%
 \begin{subfigure}{8cm}
 \centering\includegraphics[width=8cm]{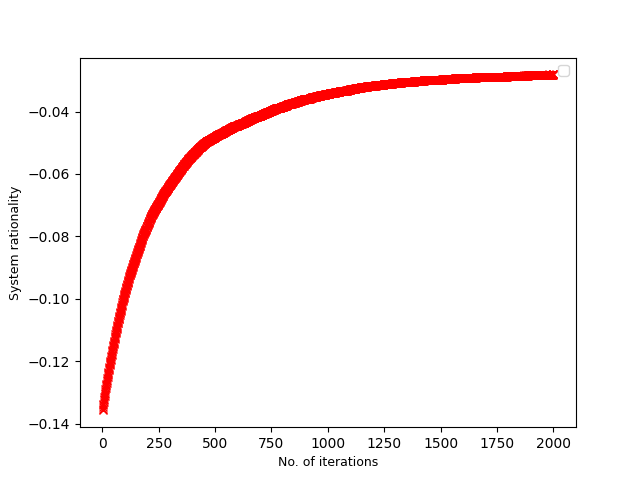} 
 \caption{Node rationality negatively \newline correlated with degree}
 \end{subfigure}\vspace{10pt}
 \caption{System rationality against the no. of iterations of network rewiring. The system rationality increases and stabilizes over $2000$ iterations of rewiring,  when the node rationality is both positively and negatively correlated with the degree.} 
 \label{sys-rationality}
\end{figure*}%

 \begin{figure*}
 \centering
 \begin{subfigure}{8cm}
 \centering\includegraphics[width=8cm]{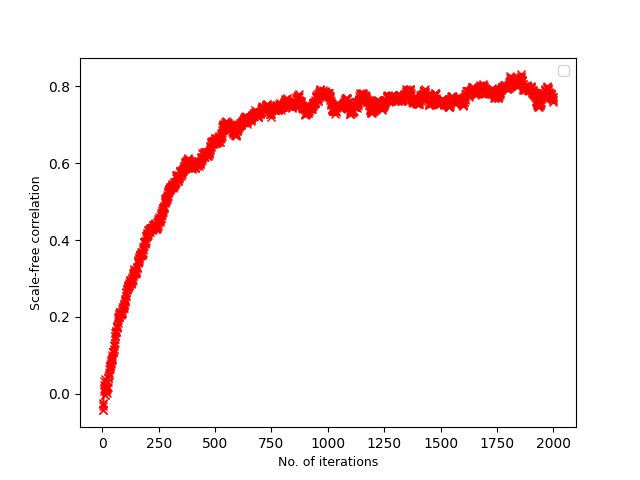}
 \caption{Node rationality positively \newline correlated with degree}
 \end{subfigure}%
 \begin{subfigure}{8cm}
 \centering\includegraphics[width=8cm]{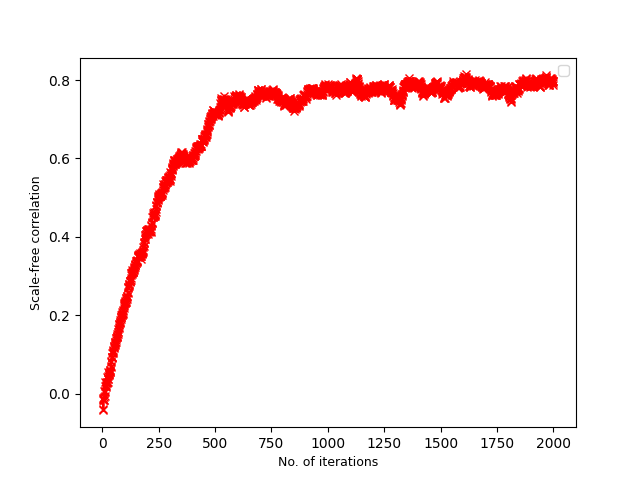}
 \caption{Node rationality negatively \newline correlated with degree}
 \end{subfigure}\vspace{10pt}
 \caption{Scale-free correlation against the no. of iterations of rewiring. The figures indicate that the scale-freeness of the topology increases over $2000$ iterations of rewiring, when the node rationality is both positively and negatively correlated with the degree.}   
 \label{correlation}
\end{figure*}%

\begin{figure*}
 \centering
 \begin{subfigure}{8cm}
 \centering\includegraphics[width=8cm]{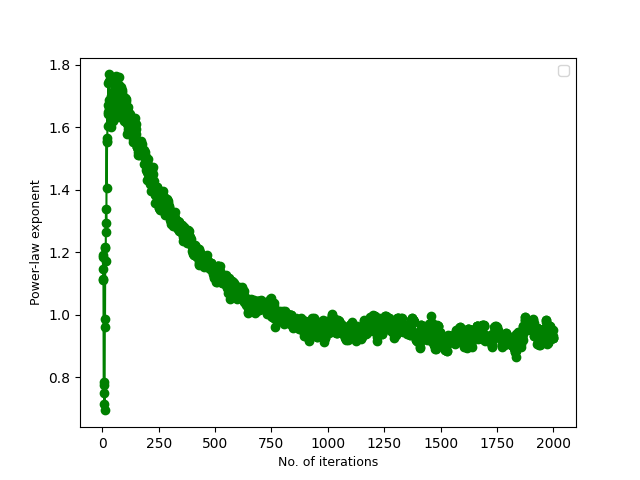}
 \caption{Node rationality positively \newline correlated with degree}
 \end{subfigure}%
 \begin{subfigure}{8cm}
 \centering\includegraphics[width=8cm]{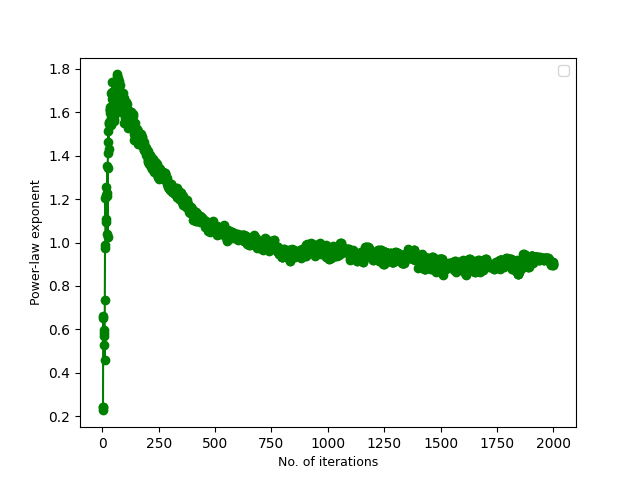} 
 \caption{Node rationality negatively \newline correlated with degree}
 \end{subfigure}\vspace{10pt}
 \caption{Power-law exponent against the no. of iterations of rewiring. Power-law exponent reaches $1$, indicative of a core-periphery topology emerging, when the node rationality is both positively and negatively correlated with the degree.}
 \label{power-law}
\end{figure*}%

\begin{figure*}
 \centering
 \begin{subfigure}{8cm}
 \centering\includegraphics[width=8cm]{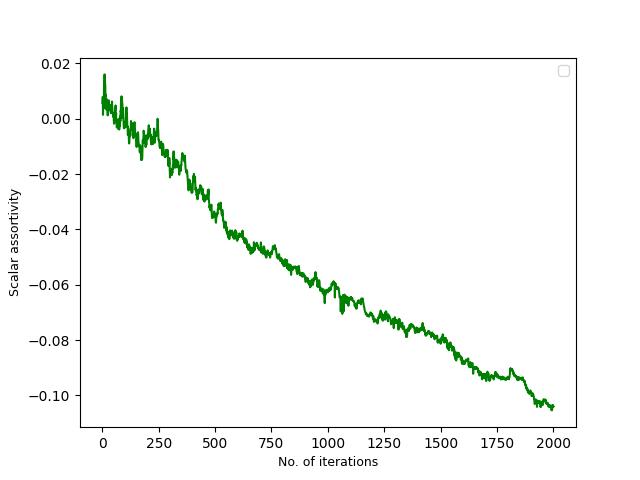}
 \caption{Node rationality positively \newline correlated with degree}
 \end{subfigure}%
 \begin{subfigure}{8cm}
 \centering\includegraphics[width=8cm]{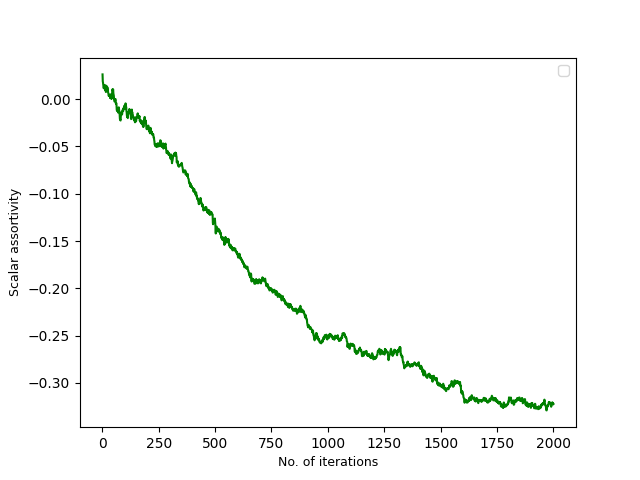}
 \caption{Node rationality negatively \newline correlated with degree}
 \end{subfigure}\vspace{10pt}
 \caption{Node rationality-based scalar-assortativity against the no. of iterations of rewiring. The scalar disassortativity increases over $2000$ iterations when the node rationality is both positively and negatively correlated with the degree.}
 \label{sclar-assor}   
\end{figure*}%

\begin{figure*}
 \centering
 \begin{subfigure}{8cm}
 \centering\includegraphics[width=8cm]{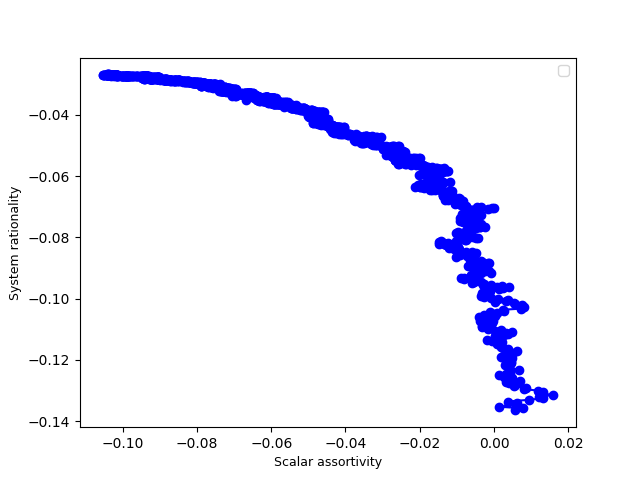}
 \caption{Node rationality positively \newline correlated with degree}
 \end{subfigure}%
 \begin{subfigure}{8cm}
 \centering\includegraphics[width=8cm]{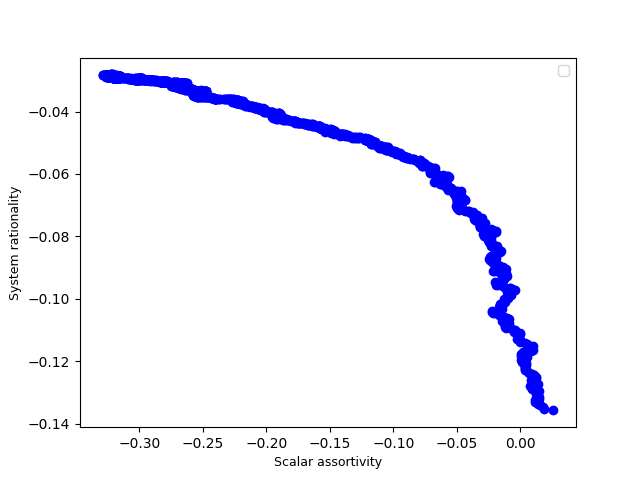}
 \caption{Node rationality negatively \newline correlated with degree}
 \end{subfigure}\vspace{10pt}
 \caption{System rationality against node rationality-based scalar-assortativity, derived from the intermediate networks obtained when the random topology is rewired to improve the system rationality.}
 \label{assor_rationality}
\end{figure*}%

Fig. \ref{sys-rationality} denotes the evolution of system rationality over multiple iterations of rewiring. The system rationality increases and stabilizes over time. The evolution of the system rationality is similar, irrespective of whether the node rationality is positively or negatively correlated with the node degree. This is to be expected as the simulated annealing based optimization process was set-up with the intention of incrementally improving the system rationality. 

The Fig. \ref{correlation} demonstrates that the network topology increases its scale-freeness when the system is evolved in such a way that the system rationality is incrementally improved. This tendency is valid when the node rationality is both positively and negatively correlated with the node degree. Further, Fig. \ref{power-law} indicates that the power-law exponent reaches the value $1$ when the network is evolved to incrementally improve the system rationality. The resultant topology is indicative of a core-periphery topology as the power-law-exponent observed is similar to that of a core-periphery topology\cite{perera2018topological}.

Fig. \ref{sclar-assor} denotes the evolution of the node rationality-based scalar-assortativity, while the system rationality is iteratively increased. It shows that the node rationality-based scalar disassortativity increases while the system rationality is incrementally increased over multiple iterations. It is important to note that this pattern holds when the node rationality is both positively and negatively correlated to node degree. This indicates that irrespective of whether the nodes with higher rationality are placed as hubs or leaf nodes, system rationality increases as long as nodes with contrasting rationality values, suggested by scalar disassortative mixing, interact with each other. 

This observation is further reinforced by Fig. \ref{assor_rationality} that depicts the system rationality against scalar-assortativity, when node rationality is both positively and negatively correlated to node degree. It shows a clear pattern that in intermediary networks, the system rationality seem to be negatively correlated to node rationality based scalar-assortativity. These results suggest that when the node rationality is distributed in a more disassortative manner, the system rationality seems to improve, irrespective of whether it's the hubs or the leaf nodes that tend to be the more rational nodes in the network.

\section{Discussion} \label{s-Conclusions}

Based on the results depicted in the previous section, it is apparent that an initially random network topology evolves into a core-periphery topology, when the system rationality is optimized. It has to be noted that this pattern holds when the node rationality both positively and negatively correlated with the node degree, suggesting that hubs need not be more rational in order to maximize system rationality, contrary to the previous work done on networks of players with topologically distributed node rationality values. 

In other words, the core-periphery topology may be just a special case of a more general network property of rationality-based scalar disassortativity that optimizes the system rationality, as core-periphery networks have been observed to have higher values of degree disassortativity~\cite{new2000,Newman2003,newman2003mixing}. Therefore, we may argue that the core-periphery networks help optimise system rationality, as demonstrated by Roman et. al ~\cite{roman2017topology}, not necessarily because of the intrinsic topology of the core-periphery networks, but due to the node rationality-based scalar disassortativity that emerges when the node rationality values are assigned to be correlated with the node degree in core-periphery networks.


 
This result is further reinforced by the fact that the scalar disassoratativity and system rationality seem to be correlated when node rationality is both positively and negatively correlated with the degree. Therefore, we argue that the system rationality may not be optimized based on the degree distribution alone, and that assortative mixing of players too is a decisive factor. From the results, it may be argued that it's the node rationality-based scalar-assortativity that drives system rationality, and not necessarily the degree distribution. The effect of network topology may play a role, as long as it facilitates more disassortative mixing of bounded node rationality.  


While the simulations in this work were conducted based on the prisoner's dilemma game, the observed negative correlation between scalar assortativity and system rationality may be valid for other games as well, such as the public goods game and the stag-hunt game. The reason is that when the rationality of interactions improve in any strategic decision-making environment, more rational behavior may emerge, thereby pushing all interactions towards Nash equilibria. Thus, node rationality-based scalar disassortativity of a network may facilitate more rational interactions, irrespective of the strategic game being adopted by the agents.


\begin{thebibliography}{41}%
\makeatletter
\providecommand \@ifxundefined [1]{%
 \@ifx{#1\undefined}
}%
\providecommand \@ifnum [1]{%
 \ifnum #1\expandafter \@firstoftwo
 \else \expandafter \@secondoftwo
 \fi
}%
\providecommand \@ifx [1]{%
 \ifx #1\expandafter \@firstoftwo
 \else \expandafter \@secondoftwo
 \fi
}%
\providecommand \natexlab [1]{#1}%
\providecommand \enquote  [1]{``#1''}%
\providecommand \bibnamefont  [1]{#1}%
\providecommand \bibfnamefont [1]{#1}%
\providecommand \citenamefont [1]{#1}%
\providecommand \href@noop [0]{\@secondoftwo}%
\providecommand \href [0]{\begingroup \@sanitize@url \@href}%
\providecommand \@href[1]{\@@startlink{#1}\@@href}%
\providecommand \@@href[1]{\endgroup#1\@@endlink}%
\providecommand \@sanitize@url [0]{\catcode `\\12\catcode `\$12\catcode
  `\&12\catcode `\#12\catcode `\^12\catcode `\_12\catcode `\%12\relax}%
\providecommand \@@startlink[1]{}%
\providecommand \@@endlink[0]{}%
\providecommand \url  [0]{\begingroup\@sanitize@url \@url }%
\providecommand \@url [1]{\endgroup\@href {#1}{\urlprefix }}%
\providecommand \urlprefix  [0]{URL }%
\providecommand \Eprint [0]{\href }%
\providecommand \doibase [0]{https://doi.org/}%
\providecommand \selectlanguage [0]{\@gobble}%
\providecommand \bibinfo  [0]{\@secondoftwo}%
\providecommand \bibfield  [0]{\@secondoftwo}%
\providecommand \translation [1]{[#1]}%
\providecommand \BibitemOpen [0]{}%
\providecommand \bibitemStop [0]{}%
\providecommand \bibitemNoStop [0]{.\EOS\space}%
\providecommand \EOS [0]{\spacefactor3000\relax}%
\providecommand \BibitemShut  [1]{\csname bibitem#1\endcsname}%
\let\auto@bib@innerbib\@empty
\bibitem [{\citenamefont {Kasthurirathna}\ and\ \citenamefont
  {Piraveenan}(2015)}]{kasthurirathna2015emergence}%
  \BibitemOpen
  \bibfield  {author} {\bibinfo {author} {\bibfnamefont {D.}~\bibnamefont
  {Kasthurirathna}}\ and\ \bibinfo {author} {\bibfnamefont {M.}~\bibnamefont
  {Piraveenan}},\ }\bibfield  {title} {\bibinfo {title} {Emergence of
  scale-free characteristics in socio-ecological systems with bounded
  rationality},\ }\href@noop {} {\bibfield  {journal} {\bibinfo  {journal}
  {Scientific reports}\ }\textbf {\bibinfo {volume} {5}},\ \bibinfo {pages}
  {10448} (\bibinfo {year} {2015})}\BibitemShut {NoStop}%
\bibitem [{\citenamefont {Roman}\ and\ \citenamefont
  {Brede}(2017)}]{roman2017topology}%
  \BibitemOpen
  \bibfield  {author} {\bibinfo {author} {\bibfnamefont {S.}~\bibnamefont
  {Roman}}\ and\ \bibinfo {author} {\bibfnamefont {M.}~\bibnamefont {Brede}},\
  }\bibfield  {title} {\bibinfo {title} {Topology-dependent rationality and
  quantal response equilibria in structured populations},\ }\href@noop {}
  {\bibfield  {journal} {\bibinfo  {journal} {Physical Review E}\ }\textbf
  {\bibinfo {volume} {95}},\ \bibinfo {pages} {052310} (\bibinfo {year}
  {2017})}\BibitemShut {NoStop}%
\bibitem [{\citenamefont {Rombach}\ \emph {et~al.}(2014)\citenamefont
  {Rombach}, \citenamefont {Porter}, \citenamefont {Fowler},\ and\
  \citenamefont {Mucha}}]{rombach2014core}%
  \BibitemOpen
  \bibfield  {author} {\bibinfo {author} {\bibfnamefont {M.~P.}\ \bibnamefont
  {Rombach}}, \bibinfo {author} {\bibfnamefont {M.~A.}\ \bibnamefont {Porter}},
  \bibinfo {author} {\bibfnamefont {J.~H.}\ \bibnamefont {Fowler}},\ and\
  \bibinfo {author} {\bibfnamefont {P.~J.}\ \bibnamefont {Mucha}},\ }\bibfield
  {title} {\bibinfo {title} {Core-periphery structure in networks},\
  }\href@noop {} {\bibfield  {journal} {\bibinfo  {journal} {SIAM Journal on
  Applied mathematics}\ }\textbf {\bibinfo {volume} {74}},\ \bibinfo {pages}
  {167} (\bibinfo {year} {2014})}\BibitemShut {NoStop}%
\bibitem [{\citenamefont {Law}\ \emph {et~al.}(2019)\citenamefont {Law},
  \citenamefont {Kasthurirathna},\ and\ \citenamefont
  {Piraveenan}}]{law2019placement}%
  \BibitemOpen
  \bibfield  {author} {\bibinfo {author} {\bibfnamefont {S.~Y.}\ \bibnamefont
  {Law}}, \bibinfo {author} {\bibfnamefont {D.}~\bibnamefont
  {Kasthurirathna}},\ and\ \bibinfo {author} {\bibfnamefont {M.}~\bibnamefont
  {Piraveenan}},\ }\bibfield  {title} {\bibinfo {title} {Placement matters in
  making good decisions sooner: the influence of topology in reaching public
  utility thresholds},\ }in\ \href@noop {} {\emph {\bibinfo {booktitle}
  {Proceedings of the 2019 IEEE/ACM International Conference on Advances in
  Social Networks Analysis and Mining}}}\ (\bibinfo {year} {2019})\ pp.\
  \bibinfo {pages} {787--795}\BibitemShut {NoStop}%
\bibitem [{\citenamefont {Newman}(2003{\natexlab{a}})}]{newman2003structure}%
  \BibitemOpen
  \bibfield  {author} {\bibinfo {author} {\bibfnamefont {M.~E.}\ \bibnamefont
  {Newman}},\ }\bibfield  {title} {\bibinfo {title} {The structure and function
  of complex networks},\ }\href@noop {} {\bibfield  {journal} {\bibinfo
  {journal} {SIAM review}\ }\textbf {\bibinfo {volume} {45}},\ \bibinfo {pages}
  {167} (\bibinfo {year} {2003}{\natexlab{a}})}\BibitemShut {NoStop}%
\bibitem [{\citenamefont {Albert}\ and\ \citenamefont
  {Barab{\'a}si}(2002)}]{albert2002statistical}%
  \BibitemOpen
  \bibfield  {author} {\bibinfo {author} {\bibfnamefont {R.}~\bibnamefont
  {Albert}}\ and\ \bibinfo {author} {\bibfnamefont {A.-L.}\ \bibnamefont
  {Barab{\'a}si}},\ }\bibfield  {title} {\bibinfo {title} {Statistical
  mechanics of complex networks},\ }\href@noop {} {\bibfield  {journal}
  {\bibinfo  {journal} {Reviews of modern physics}\ }\textbf {\bibinfo {volume}
  {74}},\ \bibinfo {pages} {47} (\bibinfo {year} {2002})}\BibitemShut {NoStop}%
\bibitem [{\citenamefont {Chung}\ \emph {et~al.}(2014)\citenamefont {Chung},
  \citenamefont {Piraveen},\ and\ \citenamefont {Hossain}}]{chung2014topology}%
  \BibitemOpen
  \bibfield  {author} {\bibinfo {author} {\bibfnamefont {K.}~\bibnamefont
  {Chung}}, \bibinfo {author} {\bibfnamefont {M.}~\bibnamefont {Piraveen}},\
  and\ \bibinfo {author} {\bibfnamefont {L.}~\bibnamefont {Hossain}},\
  }\bibfield  {title} {\bibinfo {title} {Topology of online social networks},\
  }\href@noop {} {\bibfield  {journal} {\bibinfo  {journal} {Encyclopedia of
  Social Network Analysis and Mining}\ } (\bibinfo {year} {2014})}\BibitemShut
  {NoStop}%
\bibitem [{\citenamefont {Weibull}(1997)}]{weibull1997evolutionary}%
  \BibitemOpen
  \bibfield  {author} {\bibinfo {author} {\bibfnamefont {J.~W.}\ \bibnamefont
  {Weibull}},\ }\href@noop {} {\emph {\bibinfo {title} {Evolutionary game
  theory}}}\ (\bibinfo  {publisher} {MIT press},\ \bibinfo {year}
  {1997})\BibitemShut {NoStop}%
\bibitem [{\citenamefont {Taylor}\ and\ \citenamefont
  {Jonker}(1978)}]{taylor1978evolutionary}%
  \BibitemOpen
  \bibfield  {author} {\bibinfo {author} {\bibfnamefont {P.~D.}\ \bibnamefont
  {Taylor}}\ and\ \bibinfo {author} {\bibfnamefont {L.~B.}\ \bibnamefont
  {Jonker}},\ }\bibfield  {title} {\bibinfo {title} {Evolutionary stable
  strategies and game dynamics},\ }\href@noop {} {\bibfield  {journal}
  {\bibinfo  {journal} {Mathematical biosciences}\ }\textbf {\bibinfo {volume}
  {40}},\ \bibinfo {pages} {145} (\bibinfo {year} {1978})}\BibitemShut
  {NoStop}%
\bibitem [{\citenamefont {Santos}\ \emph {et~al.}(2006)\citenamefont {Santos},
  \citenamefont {Rodrigues},\ and\ \citenamefont {Pacheco}}]{santos2006graph}%
  \BibitemOpen
  \bibfield  {author} {\bibinfo {author} {\bibfnamefont {F.~C.}\ \bibnamefont
  {Santos}}, \bibinfo {author} {\bibfnamefont {J.}~\bibnamefont {Rodrigues}},\
  and\ \bibinfo {author} {\bibfnamefont {J.}~\bibnamefont {Pacheco}},\
  }\bibfield  {title} {\bibinfo {title} {Graph topology plays a determinant
  role in the evolution of cooperation},\ }\href@noop {} {\bibfield  {journal}
  {\bibinfo  {journal} {Proceedings of the Royal Society B: Biological
  Sciences}\ }\textbf {\bibinfo {volume} {273}},\ \bibinfo {pages} {51}
  (\bibinfo {year} {2006})}\BibitemShut {NoStop}%
\bibitem [{\citenamefont {Kasthurirathna}\ \emph {et~al.}(2015)\citenamefont
  {Kasthurirathna}, \citenamefont {Piraveenan},\ and\ \citenamefont
  {Uddin}}]{kasthurirathna2015evolutionary}%
  \BibitemOpen
  \bibfield  {author} {\bibinfo {author} {\bibfnamefont {D.}~\bibnamefont
  {Kasthurirathna}}, \bibinfo {author} {\bibfnamefont {M.}~\bibnamefont
  {Piraveenan}},\ and\ \bibinfo {author} {\bibfnamefont {S.}~\bibnamefont
  {Uddin}},\ }\bibfield  {title} {\bibinfo {title} {Evolutionary stable
  strategies in networked games: the influence of topology},\ }\href@noop {}
  {\bibfield  {journal} {\bibinfo  {journal} {Journal of Artificial
  Intelligence and Soft Computing Research}\ }\textbf {\bibinfo {volume} {5}},\
  \bibinfo {pages} {83} (\bibinfo {year} {2015})}\BibitemShut {NoStop}%
\bibitem [{\citenamefont {Kasthurirathna}\ \emph {et~al.}(2014)\citenamefont
  {Kasthurirathna}, \citenamefont {Piraveenan},\ and\ \citenamefont
  {Harr{\'e}}}]{kasthurirathna2014influence}%
  \BibitemOpen
  \bibfield  {author} {\bibinfo {author} {\bibfnamefont {D.}~\bibnamefont
  {Kasthurirathna}}, \bibinfo {author} {\bibfnamefont {M.}~\bibnamefont
  {Piraveenan}},\ and\ \bibinfo {author} {\bibfnamefont {M.}~\bibnamefont
  {Harr{\'e}}},\ }\bibfield  {title} {\bibinfo {title} {Influence of topology
  in the evolution of coordination in complex networks under information
  diffusion constraints},\ }\href@noop {} {\bibfield  {journal} {\bibinfo
  {journal} {The European Physical Journal B}\ }\textbf {\bibinfo {volume}
  {87}},\ \bibinfo {pages} {1} (\bibinfo {year} {2014})}\BibitemShut {NoStop}%
\bibitem [{\citenamefont {Devaine}\ \emph {et~al.}(2014)\citenamefont
  {Devaine}, \citenamefont {Hollard},\ and\ \citenamefont
  {Daunizeau}}]{devaine2014theory}%
  \BibitemOpen
  \bibfield  {author} {\bibinfo {author} {\bibfnamefont {M.}~\bibnamefont
  {Devaine}}, \bibinfo {author} {\bibfnamefont {G.}~\bibnamefont {Hollard}},\
  and\ \bibinfo {author} {\bibfnamefont {J.}~\bibnamefont {Daunizeau}},\
  }\bibfield  {title} {\bibinfo {title} {Theory of mind: Did evolution fool
  us?},\ }\href@noop {} {\bibfield  {journal} {\bibinfo  {journal} {PloS One}\
  }\textbf {\bibinfo {volume} {9}},\ \bibinfo {pages} {e87619} (\bibinfo {year}
  {2014})}\BibitemShut {NoStop}%
\bibitem [{\citenamefont {Anh}\ \emph {et~al.}(2011)\citenamefont {Anh},
  \citenamefont {Moniz~Pereira},\ and\ \citenamefont
  {Santos}}]{han2011intention}%
  \BibitemOpen
  \bibfield  {author} {\bibinfo {author} {\bibfnamefont {H.~T.}\ \bibnamefont
  {Anh}}, \bibinfo {author} {\bibfnamefont {L.}~\bibnamefont {Moniz~Pereira}},\
  and\ \bibinfo {author} {\bibfnamefont {F.~C.}\ \bibnamefont {Santos}},\
  }\bibfield  {title} {\bibinfo {title} {Intention recognition promotes the
  emergence of cooperation},\ }\href@noop {} {\bibfield  {journal} {\bibinfo
  {journal} {Adaptive Behavior}\ }\textbf {\bibinfo {volume} {19}},\ \bibinfo
  {pages} {264} (\bibinfo {year} {2011})}\BibitemShut {NoStop}%
\bibitem [{\citenamefont {McNally}\ \emph {et~al.}(2012)\citenamefont
  {McNally}, \citenamefont {Brown},\ and\ \citenamefont
  {Jackson}}]{mcnally2012cooperation}%
  \BibitemOpen
  \bibfield  {author} {\bibinfo {author} {\bibfnamefont {L.}~\bibnamefont
  {McNally}}, \bibinfo {author} {\bibfnamefont {S.~P.}\ \bibnamefont {Brown}},\
  and\ \bibinfo {author} {\bibfnamefont {A.~L.}\ \bibnamefont {Jackson}},\
  }\bibfield  {title} {\bibinfo {title} {Cooperation and the evolution of
  intelligence},\ }\href@noop {} {\bibfield  {journal} {\bibinfo  {journal}
  {Proceedings of the Royal Society B: Biological Sciences}\ }\textbf {\bibinfo
  {volume} {279}},\ \bibinfo {pages} {3027} (\bibinfo {year}
  {2012})}\BibitemShut {NoStop}%
\bibitem [{\citenamefont {Rong}\ \emph {et~al.}(2007)\citenamefont {Rong},
  \citenamefont {Li},\ and\ \citenamefont {Wang}}]{rong2007roles}%
  \BibitemOpen
  \bibfield  {author} {\bibinfo {author} {\bibfnamefont {Z.}~\bibnamefont
  {Rong}}, \bibinfo {author} {\bibfnamefont {X.}~\bibnamefont {Li}},\ and\
  \bibinfo {author} {\bibfnamefont {X.}~\bibnamefont {Wang}},\ }\bibfield
  {title} {\bibinfo {title} {Roles of mixing patterns in cooperation on a
  scale-free networked game},\ }\href@noop {} {\bibfield  {journal} {\bibinfo
  {journal} {Physical Review E}\ }\textbf {\bibinfo {volume} {76}},\ \bibinfo
  {pages} {027101} (\bibinfo {year} {2007})}\BibitemShut {NoStop}%
\bibitem [{\citenamefont {Rong}\ and\ \citenamefont
  {Wu}(2009)}]{rong2009effect}%
  \BibitemOpen
  \bibfield  {author} {\bibinfo {author} {\bibfnamefont {Z.}~\bibnamefont
  {Rong}}\ and\ \bibinfo {author} {\bibfnamefont {Z.-X.}\ \bibnamefont {Wu}},\
  }\bibfield  {title} {\bibinfo {title} {Effect of the degree correlation in
  public goods game on scale-free networks},\ }\href@noop {} {\bibfield
  {journal} {\bibinfo  {journal} {EPL (Europhysics Letters)}\ }\textbf
  {\bibinfo {volume} {87}},\ \bibinfo {pages} {30001} (\bibinfo {year}
  {2009})}\BibitemShut {NoStop}%
\bibitem [{\citenamefont {Axelrod}(1980)}]{axelrod1980effective}%
  \BibitemOpen
  \bibfield  {author} {\bibinfo {author} {\bibfnamefont {R.}~\bibnamefont
  {Axelrod}},\ }\bibfield  {title} {\bibinfo {title} {Effective choice in the
  prisoner's dilemma},\ }\href@noop {} {\bibfield  {journal} {\bibinfo
  {journal} {Journal of conflict resolution}\ }\textbf {\bibinfo {volume}
  {24}},\ \bibinfo {pages} {3} (\bibinfo {year} {1980})}\BibitemShut {NoStop}%
\bibitem [{\citenamefont {Myerson}(1999)}]{myerson1999nash}%
  \BibitemOpen
  \bibfield  {author} {\bibinfo {author} {\bibfnamefont {R.~B.}\ \bibnamefont
  {Myerson}},\ }\bibfield  {title} {\bibinfo {title} {Nash equilibrium and the
  history of economic theory},\ }\href@noop {} {\bibfield  {journal} {\bibinfo
  {journal} {Journal of Economic Literature}\ }\textbf {\bibinfo {volume}
  {37}},\ \bibinfo {pages} {1067} (\bibinfo {year} {1999})}\BibitemShut
  {NoStop}%
\bibitem [{\citenamefont {Conlisk}(1996)}]{conlisk1996bounded}%
  \BibitemOpen
  \bibfield  {author} {\bibinfo {author} {\bibfnamefont {J.}~\bibnamefont
  {Conlisk}},\ }\bibfield  {title} {\bibinfo {title} {Why bounded
  rationality?},\ }\href@noop {} {\bibfield  {journal} {\bibinfo  {journal}
  {Journal of economic literature}\ }\textbf {\bibinfo {volume} {34}},\
  \bibinfo {pages} {669} (\bibinfo {year} {1996})}\BibitemShut {NoStop}%
\bibitem [{\citenamefont {McKelvey}\ and\ \citenamefont
  {Palfrey}(1995)}]{mckelvey1995quantal}%
  \BibitemOpen
  \bibfield  {author} {\bibinfo {author} {\bibfnamefont {R.~D.}\ \bibnamefont
  {McKelvey}}\ and\ \bibinfo {author} {\bibfnamefont {T.~R.}\ \bibnamefont
  {Palfrey}},\ }\bibfield  {title} {\bibinfo {title} {Quantal response
  equilibria for normal form games},\ }\href@noop {} {\bibfield  {journal}
  {\bibinfo  {journal} {Games and economic behavior}\ }\textbf {\bibinfo
  {volume} {10}},\ \bibinfo {pages} {6} (\bibinfo {year} {1995})}\BibitemShut
  {NoStop}%
\bibitem [{\citenamefont {Goeree}\ \emph {et~al.}(2010)\citenamefont {Goeree},
  \citenamefont {Holt},\ and\ \citenamefont {Palfrey}}]{goeree2010quantal}%
  \BibitemOpen
  \bibfield  {author} {\bibinfo {author} {\bibfnamefont {J.~K.}\ \bibnamefont
  {Goeree}}, \bibinfo {author} {\bibfnamefont {C.~A.}\ \bibnamefont {Holt}},\
  and\ \bibinfo {author} {\bibfnamefont {T.~R.}\ \bibnamefont {Palfrey}},\
  }\bibfield  {title} {\bibinfo {title} {Quantal response equilibria},\ }in\
  \href@noop {} {\emph {\bibinfo {booktitle} {Behavioural and Experimental
  Economics}}}\ (\bibinfo  {publisher} {Springer},\ \bibinfo {year} {2010})\
  pp.\ \bibinfo {pages} {234--242}\BibitemShut {NoStop}%
\bibitem [{\citenamefont {Wolpert}(2006)}]{wolpert2006information}%
  \BibitemOpen
  \bibfield  {author} {\bibinfo {author} {\bibfnamefont {D.~H.}\ \bibnamefont
  {Wolpert}},\ }\bibfield  {title} {\bibinfo {title} {Information theory?the
  bridge connecting bounded rational game theory and statistical physics},\
  }in\ \href@noop {} {\emph {\bibinfo {booktitle} {Complex Engineered
  Systems}}}\ (\bibinfo  {publisher} {Springer},\ \bibinfo {year} {2006})\ pp.\
  \bibinfo {pages} {262--290}\BibitemShut {NoStop}%
\bibitem [{\citenamefont {Camerer}\ \emph {et~al.}(2004)\citenamefont
  {Camerer}, \citenamefont {Ho},\ and\ \citenamefont
  {Chong}}]{camerer2004cognitive}%
  \BibitemOpen
  \bibfield  {author} {\bibinfo {author} {\bibfnamefont {C.~F.}\ \bibnamefont
  {Camerer}}, \bibinfo {author} {\bibfnamefont {T.-H.}\ \bibnamefont {Ho}},\
  and\ \bibinfo {author} {\bibfnamefont {J.-K.}\ \bibnamefont {Chong}},\
  }\bibfield  {title} {\bibinfo {title} {A cognitive hierarchy model of
  games},\ }\href@noop {} {\bibfield  {journal} {\bibinfo  {journal} {The
  Quarterly Journal of Economics}\ ,\ \bibinfo {pages} {861}} (\bibinfo {year}
  {2004})}\BibitemShut {NoStop}%
\bibitem [{\citenamefont {Rogers}\ \emph {et~al.}(2009)\citenamefont {Rogers},
  \citenamefont {Palfrey},\ and\ \citenamefont
  {Camerer}}]{rogers2009heterogeneous}%
  \BibitemOpen
  \bibfield  {author} {\bibinfo {author} {\bibfnamefont {B.~W.}\ \bibnamefont
  {Rogers}}, \bibinfo {author} {\bibfnamefont {T.~R.}\ \bibnamefont
  {Palfrey}},\ and\ \bibinfo {author} {\bibfnamefont {C.~F.}\ \bibnamefont
  {Camerer}},\ }\bibfield  {title} {\bibinfo {title} {Heterogeneous quantal
  response equilibrium and cognitive hierarchies},\ }\href@noop {} {\bibfield
  {journal} {\bibinfo  {journal} {Journal of Economic Theory}\ }\textbf
  {\bibinfo {volume} {144}},\ \bibinfo {pages} {1440} (\bibinfo {year}
  {2009})}\BibitemShut {NoStop}%
\bibitem [{\citenamefont {Golman}(2012)}]{golman2012homogeneity}%
  \BibitemOpen
  \bibfield  {author} {\bibinfo {author} {\bibfnamefont {R.}~\bibnamefont
  {Golman}},\ }\bibfield  {title} {\bibinfo {title} {Homogeneity bias in models
  of discrete choice with bounded rationality},\ }\href@noop {} {\bibfield
  {journal} {\bibinfo  {journal} {Journal of Economic Behavior \&
  Organization}\ }\textbf {\bibinfo {volume} {82}},\ \bibinfo {pages} {1}
  (\bibinfo {year} {2012})}\BibitemShut {NoStop}%
\bibitem [{\citenamefont {Roman}(2018)}]{roman2018dynamic}%
  \BibitemOpen
  \bibfield  {author} {\bibinfo {author} {\bibfnamefont {S.}~\bibnamefont
  {Roman}},\ }\emph {\bibinfo {title} {Dynamic and Game Theoretic Modelling of
  Societal Growth, Structure and Collapse}},\ \href@noop {} {Ph.D. thesis},\
  \bibinfo  {school} {Original typescript} (\bibinfo {year} {2018})\BibitemShut
  {NoStop}%
\bibitem [{\citenamefont {Kasthurirathna}\ \emph {et~al.}(2016)\citenamefont
  {Kasthurirathna}, \citenamefont {Piraveenan},\ and\ \citenamefont
  {Uddin}}]{kasthurirathna2016modeling}%
  \BibitemOpen
  \bibfield  {author} {\bibinfo {author} {\bibfnamefont {D.}~\bibnamefont
  {Kasthurirathna}}, \bibinfo {author} {\bibfnamefont {M.}~\bibnamefont
  {Piraveenan}},\ and\ \bibinfo {author} {\bibfnamefont {S.}~\bibnamefont
  {Uddin}},\ }\bibfield  {title} {\bibinfo {title} {Modeling networked systems
  using the topologically distributed bounded rationality framework},\
  }\href@noop {} {\bibfield  {journal} {\bibinfo  {journal} {Complexity}\
  }\textbf {\bibinfo {volume} {21}},\ \bibinfo {pages} {123} (\bibinfo {year}
  {2016})}\BibitemShut {NoStop}%
\bibitem [{\citenamefont {Gunawardana}\ \emph {et~al.}(2019)\citenamefont
  {Gunawardana}, \citenamefont {Ratnayake}, \citenamefont {Piraveenan},\ and\
  \citenamefont {Kasthurirathna}}]{gunawardana2019information}%
  \BibitemOpen
  \bibfield  {author} {\bibinfo {author} {\bibfnamefont {L.}~\bibnamefont
  {Gunawardana}}, \bibinfo {author} {\bibfnamefont {P.}~\bibnamefont
  {Ratnayake}}, \bibinfo {author} {\bibfnamefont {M.}~\bibnamefont
  {Piraveenan}},\ and\ \bibinfo {author} {\bibfnamefont {D.}~\bibnamefont
  {Kasthurirathna}},\ }\bibfield  {title} {\bibinfo {title} {Information
  theoretic approach for modeling bounded rationality in networked games},\
  }in\ \href@noop {} {\emph {\bibinfo {booktitle} {2019 IEEE Symposium Series
  on Computational Intelligence (SSCI)}}}\ (\bibinfo {organization} {IEEE},\
  \bibinfo {year} {2019})\ pp.\ \bibinfo {pages} {2100--2107}\BibitemShut
  {NoStop}%
\bibitem [{\citenamefont {Goeree}\ \emph {et~al.}(2008)\citenamefont {Goeree},
  \citenamefont {Holt},\ and\ \citenamefont {Palfrey}}]{goeree2008quantal}%
  \BibitemOpen
  \bibfield  {author} {\bibinfo {author} {\bibfnamefont {J.~K.}\ \bibnamefont
  {Goeree}}, \bibinfo {author} {\bibfnamefont {C.~A.}\ \bibnamefont {Holt}},\
  and\ \bibinfo {author} {\bibfnamefont {T.~R.}\ \bibnamefont {Palfrey}},\
  }\bibfield  {title} {\bibinfo {title} {Quantal response equilibrium},\
  }\href@noop {} {\bibfield  {journal} {\bibinfo  {journal} {The New Palgrave
  Dictionary of Economics. Palgrave Macmillan, Basingstoke}\ } (\bibinfo {year}
  {2008})}\BibitemShut {NoStop}%
\bibitem [{\citenamefont {Zhang}(2013)}]{zhang2013quantal}%
  \BibitemOpen
  \bibfield  {author} {\bibinfo {author} {\bibfnamefont {B.}~\bibnamefont
  {Zhang}},\ }\bibfield  {title} {\bibinfo {title} {Quantal response methods
  for equilibrium selection in normal form games},\ }\href@noop {} {\bibfield
  {journal} {\bibinfo  {journal} {Available at SSRN 2375553}\ } (\bibinfo
  {year} {2013})}\BibitemShut {NoStop}%
\bibitem [{\citenamefont {Cover}\ and\ \citenamefont
  {Thomas}(1991)}]{cover1991entropy}%
  \BibitemOpen
  \bibfield  {author} {\bibinfo {author} {\bibfnamefont {T.~M.}\ \bibnamefont
  {Cover}}\ and\ \bibinfo {author} {\bibfnamefont {J.~A.}\ \bibnamefont
  {Thomas}},\ }\bibfield  {title} {\bibinfo {title} {Entropy, relative entropy
  and mutual information},\ }\href@noop {} {\bibfield  {journal} {\bibinfo
  {journal} {Elements of Information Theory}\ ,\ \bibinfo {pages} {12}}
  (\bibinfo {year} {1991})}\BibitemShut {NoStop}%
\bibitem [{\citenamefont {Men{\'e}ndez}\ \emph {et~al.}(1997)\citenamefont
  {Men{\'e}ndez}, \citenamefont {Pardo}, \citenamefont {Pardo},\ and\
  \citenamefont {Pardo}}]{menendez1997jensen}%
  \BibitemOpen
  \bibfield  {author} {\bibinfo {author} {\bibfnamefont {M.}~\bibnamefont
  {Men{\'e}ndez}}, \bibinfo {author} {\bibfnamefont {J.}~\bibnamefont {Pardo}},
  \bibinfo {author} {\bibfnamefont {L.}~\bibnamefont {Pardo}},\ and\ \bibinfo
  {author} {\bibfnamefont {M.}~\bibnamefont {Pardo}},\ }\bibfield  {title}
  {\bibinfo {title} {The jensen-shannon divergence},\ }\href@noop {} {\bibfield
   {journal} {\bibinfo  {journal} {Journal of the Franklin Institute}\ }\textbf
  {\bibinfo {volume} {334}},\ \bibinfo {pages} {307} (\bibinfo {year}
  {1997})}\BibitemShut {NoStop}%
\bibitem [{\citenamefont {Newman}(2002)}]{newman2002assortative}%
  \BibitemOpen
  \bibfield  {author} {\bibinfo {author} {\bibfnamefont {M.~E.}\ \bibnamefont
  {Newman}},\ }\bibfield  {title} {\bibinfo {title} {Assortative mixing in
  networks},\ }\href@noop {} {\bibfield  {journal} {\bibinfo  {journal}
  {Physical review letters}\ }\textbf {\bibinfo {volume} {89}},\ \bibinfo
  {pages} {208701} (\bibinfo {year} {2002})}\BibitemShut {NoStop}%
\bibitem [{\citenamefont {Newman}(2003{\natexlab{b}})}]{newman2003mixing}%
  \BibitemOpen
  \bibfield  {author} {\bibinfo {author} {\bibfnamefont {M.~E.}\ \bibnamefont
  {Newman}},\ }\bibfield  {title} {\bibinfo {title} {Mixing patterns in
  networks},\ }\href@noop {} {\bibfield  {journal} {\bibinfo  {journal}
  {Physical review E}\ }\textbf {\bibinfo {volume} {67}},\ \bibinfo {pages}
  {026126} (\bibinfo {year} {2003}{\natexlab{b}})}\BibitemShut {NoStop}%
\bibitem [{\citenamefont {Piraveenan}\ \emph {et~al.}(2010)\citenamefont
  {Piraveenan}, \citenamefont {Prokopenko},\ and\ \citenamefont
  {Zomaya}}]{piraveenan2010assortative}%
  \BibitemOpen
  \bibfield  {author} {\bibinfo {author} {\bibfnamefont {M.}~\bibnamefont
  {Piraveenan}}, \bibinfo {author} {\bibfnamefont {M.}~\bibnamefont
  {Prokopenko}},\ and\ \bibinfo {author} {\bibfnamefont {A.}~\bibnamefont
  {Zomaya}},\ }\bibfield  {title} {\bibinfo {title} {Assortative mixing in
  directed biological networks},\ }\href@noop {} {\bibfield  {journal}
  {\bibinfo  {journal} {IEEE/ACM Transactions on computational biology and
  bioinformatics}\ }\textbf {\bibinfo {volume} {9}},\ \bibinfo {pages} {66}
  (\bibinfo {year} {2010})}\BibitemShut {NoStop}%
\bibitem [{\citenamefont {Van~Laarhoven}\ and\ \citenamefont
  {Aarts}(1987)}]{van1987simulated}%
  \BibitemOpen
  \bibfield  {author} {\bibinfo {author} {\bibfnamefont {P.~J.}\ \bibnamefont
  {Van~Laarhoven}}\ and\ \bibinfo {author} {\bibfnamefont {E.~H.}\ \bibnamefont
  {Aarts}},\ }\bibfield  {title} {\bibinfo {title} {Simulated annealing},\ }in\
  \href@noop {} {\emph {\bibinfo {booktitle} {Simulated annealing: Theory and
  applications}}}\ (\bibinfo  {publisher} {Springer},\ \bibinfo {year} {1987})\
  pp.\ \bibinfo {pages} {7--15}\BibitemShut {NoStop}%
\bibitem [{\citenamefont {Barab{\'a}si}\ and\ \citenamefont
  {Albert}(1999)}]{barabasi1999emergence}%
  \BibitemOpen
  \bibfield  {author} {\bibinfo {author} {\bibfnamefont {A.-L.}\ \bibnamefont
  {Barab{\'a}si}}\ and\ \bibinfo {author} {\bibfnamefont {R.}~\bibnamefont
  {Albert}},\ }\bibfield  {title} {\bibinfo {title} {Emergence of scaling in
  random networks},\ }\href@noop {} {\bibfield  {journal} {\bibinfo  {journal}
  {science}\ }\textbf {\bibinfo {volume} {286}},\ \bibinfo {pages} {509}
  (\bibinfo {year} {1999})}\BibitemShut {NoStop}%
\bibitem [{\citenamefont {Perera}\ \emph {et~al.}(2018)\citenamefont {Perera},
  \citenamefont {Bell}, \citenamefont {Piraveenan}, \citenamefont
  {Kasthurirathna},\ and\ \citenamefont {Parhi}}]{perera2018topological}%
  \BibitemOpen
  \bibfield  {author} {\bibinfo {author} {\bibfnamefont {S.~S.}\ \bibnamefont
  {Perera}}, \bibinfo {author} {\bibfnamefont {M.~G.}\ \bibnamefont {Bell}},
  \bibinfo {author} {\bibfnamefont {M.}~\bibnamefont {Piraveenan}}, \bibinfo
  {author} {\bibfnamefont {D.}~\bibnamefont {Kasthurirathna}},\ and\ \bibinfo
  {author} {\bibfnamefont {M.}~\bibnamefont {Parhi}},\ }\bibfield  {title}
  {\bibinfo {title} {Topological structure of manufacturing industry supply
  chain networks},\ }\href@noop {} {\bibfield  {journal} {\bibinfo  {journal}
  {Complexity}\ }\textbf {\bibinfo {volume} {2018}} (\bibinfo {year}
  {2018})}\BibitemShut {NoStop}%
\bibitem [{\citenamefont {Newman}(2000)}]{new2000}%
  \BibitemOpen
  \bibfield  {author} {\bibinfo {author} {\bibfnamefont {M.~E.~J.}\
  \bibnamefont {Newman}},\ }\bibfield  {title} {\bibinfo {title} {Models of the
  small world},\ }\href {https://doi.org/10.1023/A:1026485807148} {\bibfield
  {journal} {\bibinfo  {journal} {Journal of Statistical Physics}\ }\textbf
  {\bibinfo {volume} {101}},\ \bibinfo {pages} {819} (\bibinfo {year}
  {2000})}\BibitemShut {NoStop}%
\bibitem [{\citenamefont {Newman}(2003{\natexlab{c}})}]{Newman2003}%
  \BibitemOpen
  \bibfield  {author} {\bibinfo {author} {\bibfnamefont {M.~E.~J.}\
  \bibnamefont {Newman}},\ }\bibfield  {title} {\bibinfo {title} {Mixing
  patterns in networks},\ }\href@noop {} {\bibfield  {journal} {\bibinfo
  {journal} {Physical Review E}\ }\textbf {\bibinfo {volume} {67}},\ \bibinfo
  {pages} {026126} (\bibinfo {year} {2003}{\natexlab{c}})}\BibitemShut
  {NoStop}%
\end{thebibliography}
\end{document}